\documentstyle[10pt]{article}
\def\bfl{\begin{flushleft}}
\def\efl{\end{flushleft}}
\def\bfr{\begin{flushright}}
\def\efr{\end{flushright}}
\def\bc{\begin{center}}
\def\ec{\end{center}}
\def\be{\begin{equation}}
\def\ee{\end{equation}}
\def\ba{\begin{eqnarray}}
\def\ea{\end{eqnarray}}

\def\lb#1{\label{#1}}

\def\text#1{\mbox{#1}}
\def\drm{\text{d}}

\def\Exp#1{\,\text{e}^{#1}}

\def\Het{$^3\text{He}$ }

\def\Sin#1{\, \text{sin}\left( #1 \right) } 
\def\Cos#1{\, \text{cos}\left( #1 \right) } 
\def\Cosh#1{\, \text{cosh}\left( #1 \right) }

\def\Sech#1#2{\, \text{sch}^{#1}\left(#2 \right) }

\begin{document}
~\\
\bfl
{\Large \bf
Non-linear phenomena in electrical circuits:
Simulation of non-linear relativistic field theory and
possible applications
}
\efl

~~\\
\bc

{\large Konstantin G. Zloshchastiev}\\
~~\\
E-mail: zlosh@email.com,
URL(s): http://zloshchastiev.webjump.com, http://zloshchastiev.cjb.net

\ec

~~\\

\abstract{\large
We propose the non-accelerator non-low-temperature simulator 
of quantum-field effects
which is based on the feeder circuits with the special feedback.
By means of it one can study the field models which contain 
fundamental concepts in
the modern field theory but do not exist in nature in a separate form.
Besides, several field phenomena might find technological applications
by virtue of the electrical analogy.
}

~\\
~\\
PACS number(s):  07.50.Ek, 11.10.Lm\\

%07.50.Ek  Circuits and circuit components
%07.50.Qx  Signal processing electronics
%
%61.72.Hh  Indirect evidence of dislocations and other defects (resistivity,
%          slip, creep, strains, internal friction, EPR, NMR, etc.)
%61.72.Ji  Point defects (vacancies, interstitials, color centers, etc.) and
%          defect clusters
%
%84.30.Ng  Oscillators, pulse generators, and function generators
%84.30.Sk  Pulse and digital circuits
%
%11.10.Kk  Field theories in dimensions other than four
%11.10.Lm  Nonlinear or nonlocal theories and models
%11.27.+d  Extended classical solutions; cosmic strings, domain walls,
%          texture
%

~~\\
%\newpage
\large

Nowadays the methods of non-accelerator and non-astronomical 
investigations of the 
relativistic quantum field theory and gravity are of great interest.
The most significant progress was reached by means of the simulations 
based on the superfluid helium.
As was shown in numerous works the superfluid phases of \Het 
can simulate several phenomena in quantum field theory and gravity, 
namely, black holes, surface gravity, Hawking radiation, horizons, 
ergoregions, trapped surfaces, 
baryogenesis, vortexes, strings, textures, standard electroweak model, 
etc \cite{unr}.
Thereby, such interdisciplinary analogies might be very 
useful not only as a good tool for the verification of theoretical
conceptions and models but also as promising source of new technologies.

In present paper we propose another non-accelerator simulator working
at reasonably high temperature (unlike of superfluid helium) for 
studying and verifying of several phenomena of scalar field theory.
It will be shown that by means of it one can study the 
field models which are the basis of modern physics but
do not exist in nature separately.
This simulator is based on the standard wave phenomena in electrical
circuits.
In principle, the non-linear waves of electrical
nature are being widely studied and applied in optics where 
the (non-linear) light wave propagates inside a light guide.
Nevertheless and unlike this,
our present study will include the study of
conditions of both the electrical potential wave propagation in 
{\it optically opaque} medium and the 
wave propagation of the charge {\it current}, i.e., electrons.
These two circumstances strictly differ our case from the optical one,
especially in what concerns the technological applications.

So, 
let us consider the usual twin feeder, $U$ is the potential difference, 
$I$ is the current, $C$, $L$ and $R$ are
the specific wire-to-wire capacitance, inductance and 
resistance respectively.
Therefore,  
the capacitance, inductance and resistance of a small line section 
have to be
$\delta C = C \delta x$,
$\delta L = L \delta x$,
$\delta R = R \delta x$, 
where $\delta x$ is the length of the section.

The section's charge $\delta Q = C U \delta x$ varies because of both
the difference of currents in the points $x$ and $x + \delta x$ and
lateral leakage current through the isolation of the feeder.
Hence we have
\be                                                               \lb{e1}
\partial_t \delta Q = C \delta x \partial_t U =
I(x) - I(x+\delta x) - G U \delta x,
\ee
where $G$ is the leakage coefficient.
In the limit $\delta x \to 0$ we therefore obtain
\be                                                               \lb{e2}
C \partial_t U + \partial_x I + G U = 0.
\ee
The second circuit rule yields
\be                                                               \lb{e3}
\delta x ( R I + L c^{-2} \partial_t I) +
U(x+\delta x) - U(x) =0,
\ee
or in the limit $\delta x \to 0$: 
\be                                                               \lb{e4}
\partial_x U + L c^{-2} \partial_t I + I R = 0.
\ee
For simplicity further we will assume the capacitance, inductance and
leakage coefficient to
be constant.
Then the equality of mixed derivatives of the voltage yields the telegraph
equation for the line current $I (x,t)$ 
(an analogous expression can be obtained for $U$): 
\be                                                               \lb{e5}
\frac{1}{\tilde c^2}\partial_{tt} I - \partial_{xx} I 
+ \frac{G L}{c^2} \partial_t I + G I R + C \partial_t (I R)= 0,
\ee
where $\tilde c = c/\sqrt{L C}$ is the (effective) propagation speed of
circuit oscillations.
Formally one could suppose that this speed can be more then the speed of
light in vacuum.
However, we should not forget that the above-mentioned expressions were
obtained in the quasistationary approximation when the wave period of
electromagnetic 
field has to be much more than the time of field propagation.

Further, if one assumes the 
excellent isolation $G=0$ then (\ref{e5})
loses the term causing the exponential attenuation of current, and we
have the following wave equation:
\be                                                               \lb{e6}
(1/\tilde c^{2})\, \partial_{tt} I   - \partial_{xx} I 
+ C \partial_t (I R)= 0,
\ee
opening a wide prospect for simulation of several wave-like phenomena
in physics. 

For instance, one can pick the (generalized) conductor or feedback system
with nonlinear response of the resistance to transmitted current 
such that 
\be                                                               \lb{e7}
R = \frac{1}{C I}
\left[
      \int V'(I) \drm t + A(x)
\right],
\ee
where $A(x)$ is an arbitrary function, $V(I(x,t))$ is the set function of 
current, the prime means the derivative with respect to $I$,
the integral is meant as a primitive.
Provided the resistance chosen in such a way
the equation (\ref{e6}) has 
the form of the relativistic scalar field 
equation
\be                                                               \lb{e8}
(1/\tilde c^{2})\, \partial_{tt} I   - \partial_{xx} I 
+ V'(I) = 0,
\ee
which is the equation of motion for the self-interacting scalar field 
model described by the Lagrangian
\be                                                               \lb{e9}
L=
\frac{1}{2}
\left[
\frac{1}{\tilde c^2} (\partial_t I)^2 - (\partial_x I)^2 
\right] - V(I).
\ee
It should be noted that the consideration of the lateral leakage current
just complicates the requirement for the line resistance:
\[
\frac{C c^2}{G L} R = 
\frac{\Exp{-\frac{G}{C} t}}
     {I}
\left[
      \int
      \Exp{\frac{G}{C} t}
      \left(
             \frac{c^2 V'(I)}{G L} + \frac{G I}{C}
      \right)
      \drm t 
      + A(x)
\right]
-1,
\]
but does not change the picture in principle except the cases when 
exponential growth or damping can arise experimental problems.

Therefore, the proposed electrical circuit indeed can simulate scalar 
field (\ref{e9}) of itself that
seems to be very important because the many concepts lying in 
foundations of modern field theory (such as the spontaneous breaking of 
symmetry, topologically nontrivial solutions, instanton effects etc.) 
can be experimentally studied separately from other field phenomena.
Besides the evident opportunities concerning the verification and 
visualization of the theoretical predictions, 
by virtue of the electrical analogy one can use certain
field phenomena for practical purposes.
First of all, it applies to several topologically nontrivial solutions 
of (\ref{e8}) which in this connection mean the signals maintaining
initial form during indefinitely long time even in the presence of
dissipative factors.
Such a stability is known to be
provided by the presence of several conserving values,
first of all,
topological index (also known as the topological charge) \cite{raj}.

Let us consider the simplest $\varphi^4$ theory where
\be                                                               \lb{e10} 
V(I) = \frac{\lambda}{4}
\left(
      I^2 - \frac{m^2}{\lambda}
\right)^2,
\ee
where $\lambda$ and $m$ are positive constant parameters.
This theory admits the topological self-dual kink solution
\be                                                               \lb{e11}
I^{(k)} (x,t) = \frac{m}{\sqrt{\lambda}} 
\tanh{\frac{m \rho}{\sqrt{2}}},
\ee
where
\[
\rho = \frac{x -v t}{\sqrt{1-(v/\tilde c)^2}},
\]
and $v=\text{const}$ is the propagation velocity of the kink.
This solution has the nonzero topological charge  
\[
Q_{(t)} = \frac{\sqrt{\lambda}}{m} 
\left[
      I(+\infty,t) - I(-\infty,t) 
\right] = 2,
\]
and can be interpreted
as the relativistic (quasi) particle with the localized
``energy'' density
\[
\varepsilon (x,t) = 
\frac{1}{2}
\left[
\frac{1}{\tilde c^2} (\partial_t I)^2 + (\partial_x I)^2 
\right] + V(I),
\]
hence
\be                                                          \lb{e12}
\varepsilon^{(k)} (x,t) = 
\frac{m^4}{2 \lambda [1- (v/\tilde c)^2]}
\Sech{4}
     {
      \frac{m \rho}{\sqrt{2}} 
     },                                                        
\ee
and the conserved total ``energy'' has the form of the energy of
a massive relativistic quasi-particle:
\[
E = \int\limits_{-\infty}^{+\infty}
\varepsilon (x,t)\ \drm x =
\frac{\mu \tilde c^2}{\sqrt{1-(v/\tilde c)^2}}, \ \
\mu^{(k)} = \frac{2\sqrt{2}}{3} \frac{m^3}{\lambda},
\]
see \cite{zlo003} and references therein.

It should be noted that for experimental purposes  
it is convenient to use the solutions shifted in
such a way to remove singularities, zeros and negative values everywhere
when it is required by the physical sense of the electrical values.
For example, one can consider the shifted kink solution
\[                                                     
I^{(k)} (x,t) = \frac{m}{\sqrt{\lambda}} 
\left[
a+
\tanh{\frac{m \rho}{\sqrt{2}}} 
\right],\ \ a = \text{const}>1,
\]
thereby the potential energy (\ref{e10}) is modified insufficiently.
As for the whole Lagrangian it has to be defined always 
at least up to additive and multiplicative constants.

Another, even more interest example for simulation
is the sin-Gordon theory \cite{bems} which,
at first, has to be nonrenormalizable on the quantum level, 
and, second, admits several
nonlinear soliton solutions which are proven 
to be of two types (soliton and doublet) and preserving an initial
shape even after the interactions with each other
in spite of the superposition principle is not valid.
The potential function has the form
\be                                                               \lb{e13} 
V(I) = - \frac{m^4}{\lambda}
\left[
      \Cos{\frac{\sqrt{\lambda} I}{m}} - 1
\right]^2,
\ee
and the scattered solitons and doublets are described respectively by the 
expressions
\ba                                                               
&&I^{(s)} (x,t) = \frac{4 m}{\sqrt{\lambda}}\, 
\text{arctan}\exp{(m \rho)},                                        \lb{e14}\\
&&I^{(d)} (x,t) = \frac{4 m}{\sqrt{\lambda}}\, \text{arctan}
\left[
      \frac{
            \Sin{
                 \frac{u m t \tilde c}{\sqrt{1+u^2}}
                }
           }
           {
           u
           \Cosh{
                 \frac{m x}{\sqrt{1+u^2}}
                }
           }
\right],                                                         \lb{e15}
\ea 
where $u$ is the dimensionless parameter determining the period
of the doublet solution (\ref{e15}).
As was mentioned, besides the separate solutions (\ref{e14}) 
and (\ref{e15}) there were discovered the solutions describing the elastic 
scattering processes between $N$ solitons (see Refs. \cite{raj,akns}
and references therein).
From the viewpoint of the theory of signal systems it means that in the
circuit there can exist an arbitrary number of such nonlinear signals which 
have not to be distorted even 
when transmitting through each other and presence of dissipative effects.

\def\CMPh{Commun. Math. Phys.}
\def\JPh{J. Phys.}
\def\JPhA{J. Phys. A.: Math. Gen.}
\def\CJP{Czech. J. Phys.}
\def\FP{Fortschr. Phys.}
\def\LMPh {Lett. Math. Phys.}
\def\MPL {Mod. Phys. Lett.}
\def\NPh  {Nucl. Phys.}
\def\PhE  {Phys.Essays}
\def\PhL  {Phys. Lett.}
\def\PhR  {Phys. Rev.}
\def\PhRL {Phys. Rev. Lett.}
\def\PhRp {Phys. Rep.}
\def\NCim {Nuovo Cimento}
\def\NuPB {Nucl. Phys.}
\def\GRG {Gen. Relativ. Gravit.}
\def\CQG {Class. Quantum Grav.}
\def\prp {report}
\def\Prp {Report}
\def\And{~and~}

\def\jn#1#2#3#4#5{#5 {\it #1}{#2} {\bf #3} {#4}}   %IOP style
% #1 tittle  #2 ser  #3 vol  #4 page  #5 year
\def\boo#1#2#3#4#5{#4 {\it #1} ({#3}: {#2}){#5}}   %IOP style
% #1 tittle  #2 publisher  #3 place  #4 year  #5 page/, p 789/
\def\prpr#1#2#3#4#5{{#5}{ #3}{#4}}                 %IOP style
% #1 tittle  #2 place #3 \Prp #4No #5 year 

%\newpage

\end{document}